
\def\a{\alpha}
\def\b{\beta}
\def\g{\gamma}
\def\r{\rho}
\def\e{\eta}
\def\l{\lambda}

\def\CC{{\mathchoice
{\rm C\mkern-8mu\vrule height1.45ex depth-.05ex
width.05em\mkern9mu\kern-.05em}
{\rm C\mkern-8mu\vrule height1.45ex depth-.05ex
width.05em\mkern9mu\kern-.05em}
{\rm C\mkern-8mu\vrule height1ex depth-.07ex
width.035em\mkern9mu\kern-.035em}
{\rm C\mkern-8mu\vrule height.65ex depth-.1ex
width.025em\mkern8mu\kern-.025em}}}

\def\RR{{\rm I\kern-1.6pt {\rm R}}}

\def\ZZ{{\rm Z}\kern-3.8pt {\rm Z} \kern2pt}

\input phyzzx.tex

\def\np{Nucl. Phys.}
\def\pl{Phys. Lett.}

\def\ap{Ann. Phys.}
\def\cmp{Comm. Math. Phys.}
\def\ijmp{Int. J. Mod. Phys.}
\def\mpl{Mod. Phys. Lett.}

\def\am{Ann. of Math.}

\def\phyrep{Phys. Rep.}
\def\am{Adv. in Math.}

\tolerance=500000
\overfullrule=0pt
\Pubnum={US-FT-20/95}
\pubnum={US-FT-20/95}
\date={July, 1995}
\pubtype={}
\titlepage

\title{THE BRST SYMMETRY OF AFFINE LIE SUPERALGEBRAS AND NON-CRITICAL
STRINGS}
\author{I.P. ENNES }
\address{Departamento de F\'\i sica de
Part\'\i culas, \break Universidad de Santiago, \break
E-15706 Santiago de Compostela, Spain. \break
{\rm e-mail: ennes@gaes.usc.es} }
\author{J.M. ISIDRO}
\address{Department of Physics,\break
Brandeis University, Waltham, MA 02254, USA\break
{\rm e-mail: isidro@binah.cc.brandeis.edu}}
\author{A.V. RAMALLO }
\address{Departamento de F\'\i sica de
Part\'\i culas, \break Universidad de Santiago, \break
E-15706 Santiago de Compostela, Spain.\break
{\rm e-mail: alfonso@gaes.usc.es}}

\abstract{ The topological field theories associated with
affine Lie superalgebras are constructed. Their BRST symmetry is
characterised by a Kazama algebra containing spin 1, 2 and
3 operators and closes linearly. Under this symmetry all
operators are grouped into BRST doublets. The relation between the
models constructed and non-critical string theories is explored.}

\endpage
\pagenumber=1
\hyphenation {o-pe-ra-tor}

\chapter{Introduction.}

The study of topological quantum field theories
\REF\wittop{E. Witten \journal\cmp&117(88)353.}
\REF\bbrt{For a review see D. Birmingham, M. Blau, M. Rakowski and G. Thompson
\journal\phyrep&209(91)129.}
[\wittop,\bbrt] has aroused great interest recently.
These theories, being endowed with a BRST symmetry, are
 models with no local degrees of freedom, so
all local excitations can be eliminated once the
topological symmetry has been fixed. In two dimensions
there exist conformal field theories which, in addition, are
topological field theories. These are the so-called Topological
Conformal Field Theories (TCFT)
\REF\dij{R. Dijkgraaf, E. Verlinde and H. Verlinde
\journal\np&B352(91)59.;``Notes on topological string theory
and 2d quantum gravity", Proceedings of the Trieste spring
school 1990, edited by M.Green et al. (World Scientific,
Singapore,1991).}[\dij].

A method to generate new TCFT's
consists in studying the BRST structure of different chiral algebras
that extend the Virasoro symmetry. This method has been applied
in ref.
\REF\tca{J. M. Isidro and A. V. Ramallo\journal\pl&B316(93)488.}
[\tca] to the case of an affine Lie algebra, whereas in ref.
\REF\tkt{J. M. Isidro and A. V. Ramallo\journal\pl&B340(94)48.}
[\tkt] this analysis was extended to the case of a superconformal
current algebra. It is the purpose of this paper to explore, using
the same  procedure, the TCFT's based on  affine Lie
superalgebras.

The topological symmetry of a TCFT is encoded in its topological
algebra, which is the operator algebra closed by the chiral
algebra of the TCFT and the BRST current. It was checked in refs.
\REF\gln{J. M. Isidro and A. V. Ramallo\journal\np&B414(94)715.}
[\tca,\tkt,\gln] that the topological algebra of a TCFT
possessing a non-abelian current algebra must include operators of
dimensions one, two and three. This algebra is the so-called
Kazama algebra
\REF\kazama{Y. Kazama \journal\mpl&A6(91)1321.}[\kazama], which
differs from the standard twisted $N=2$ superconformal algebra
\REF\EY{T. Eguchi and S.-K. Yang \journal\mpl&A4(90)1653;
T. Eguchi, S. Hosono and S.-K. Yang \journal\cmp&140(91)159.}[\EY].
The former includes two dimension-three operators and
can be regarded as an extension of the latter. The extended
nature of the Kazama algebra seems to be an unavoidable
consequence of the underlying non-abelian current symmetry.

The representation of the BRST symmetry found in refs.
[\tca,\tkt,\gln] requires  the level of the current algebra to
be fixed to some critical value related to the dual Coxeter
number of the Lie algebra. When the matter sector of
the currents is realised by means of two decoupled currents, it
is only the sum of the two levels that is constrained. In this
two-current realization, the TCFT for a Lie algebra ${\cal G}$ has a
nice interpretation as a ${\cal G}/{\cal G}$ coset model
\REF\witGG{E. Witten \journal\cmp&144(92)189.}
[\witGG]. After a suitable
deformation these theories have been shown to be related to
non-critical $W$-strings
\REF\yank{M. Spiegelglas and S. Yankielowicz
\journal\np&393(93)301;  O. Aharony et al.\journal\np&B399(93)527
 \journal\pl&B289(92)309 \journal\pl&B305(93)35.}
\REF\hu{H.L. Hu and M. Yu \journal\pl&B289(92)302
\journal\np&B391(93)389.}
\REF\sadov{V. Sadov\journal\ijmp&A8(93)5115.}
[\yank,\hu,\sadov].

Lie superalgebras seem to play an important role in the
construction and classification of extended superconformal
 algebras
\REF\sevrin{A. Sevrin, W. Troost and A. van Proeyen
\journal\pl&B208(88)447.}
\REF\petersen{K. Ito, J. O . Madsen and J. L. Petersen
\journal\pl&B318(93)315.}
[\sevrin,\petersen] and supersymmetric Toda field theories.
Given a Lie superalgebra which admits
a purely fermionic simple root system, one can
construct an $N=1$ supersymmetric Toda model
\REF\komata{S. Komata, K. Mohri and H. Nohara
\journal\np&B359(91)168.}
\REF\evans{J. Evans and T. Hollowood\journal\np&B352(91)723.}
\REF\inami{T. Inami and K.-I. Izawa\journal\pl&B225(91)523.}
[\komata,\evans,\inami]. Application of the method of
 hamiltonian reduction to Lie (super)algebras
leads to (super) ${\cal W}$-algebras
\REF\bershadsky{M. Bershadsky and H. Ooguri
\journal\cmp&126(89)49\journal\pl&B229(89)374.}
\REF\schoutens{For a review see P. Bouwknegt
and K. Schoutens\journal\phyrep&223(93)183.}
[\bershadsky,\schoutens].
There is also a possibility that string theories
may be classified in terms of superalgebras. For
example, the twisted superconformal symmetry for
strings with $N-2$ supersymmetries has been
constructed in
\REF\boresch{A. Boresch, K. Landsteiner, W. Lerche and A. Sevrin,
\journal\np&B436(95)609.}
[\boresch] via the quantum
hamiltonian reduction of ${\rm osp}(N\vert 2)$.

It therefore seems desirable to extend the
formalism initiated in  [\tca,\tkt,\gln] to cover the
general case of an arbitrary Lie superalgebra. We shall show
below that the approach followed in refs. [\tca,\tkt] can be
easily generalised to TCFT's possessing an affine Lie
superalgebra. The BRST algebra for these theories is the same as
in the bosonic case, which yields a new representation of the
topological Kazama algebra. Moreover we provide arguments to
support the idea that these theories are related to non-critical
superstring theories.

This paper is organised as follows. In section 2, after
 a brief introduction to current
superalgebras and the associated Sugawara constructions
in two  dimensions, the
topological algebra is developed explicitly.  A two-current
construction is also possible, which
paves the way for the interpretation of the theory as a gauged,
supergroup-valued
Wess-Zumino-Witten (WZW) model; this is done in section 3.
In section 4 we turn our attention to the relation
of the model constructed with the non-critical superstring
theories. We shall find a suggestive connection between the
${\rm osp}(1\vert 2)$ and ${\rm osp}(2\vert 2)$ theories and
non-critical superstring theories with one and two
supersymmetries respectively. These results generalise the
relation between non-critical $W_N$-strings and ${\rm sl}(N)/{\rm sl}(N)$
coset theories [\yank,\hu,\sadov]. Finally, our work is
summarised in section 5, together with some conclusions,
comments and suggestions for future work.

\chapter{Construction of the topological algebra.}

Let ${\cal G}$ be a finite-dimensional superalgebra over
the complex field
\REF\kac{V. G. Ka\v c\journal\cmp&53(77)31 \journal\am&26(77)8.}
\REF\scheunert{M. Scheunert,
``The Theory of Lie Superalgebras",
{\sl Lect. Notes in Math.} 716,
Springer-Verlag, Berlin (1979).}
\REF\frappat{L. Frappat, A. Sciarrino and P. Sorba
\journal\cmp&121(89)457.}
[\kac,\scheunert,\frappat]. Call
${\cal G}_B$ and ${\cal G}_F$  the even
(\ie, bosonic) and odd  (\ie, fermionic)
subspaces of ${\cal G}$, spanned respectively by basis vectors
$T_a$, $a=1,2, \ldots d_B$, and $T_{\a}$, $\a =1,2,
\ldots d_F$, where $d_B$
and $d_F$ are the respective dimensions of
${\cal G}_B$ and ${\cal G}_F$.
In the
following, latin indices $a, b,\ldots$ will run from
1 to $d_B$, while greek
indices $\alpha, \beta,\ldots$ from 1 to $d_F$.
The {\it superdimension}
of ${\cal G}$
is $d_{B}-d_{F}$. Denote the (generalised) Lie bracket by $[\,,]$. One has
$$
[T_a,T_b]=f_{ab}^{c}T_c\qquad
[T_a,T_{\alpha}]=f_{a\alpha}^{\beta}T_{\beta}\qquad
[T_{\alpha},T_{\beta}]=f_{\alpha \beta}^c T_c.
\eqn\za
$$
The generalised Lie bracket satisfies a graded Jacobi identity
$$
(-1)^{g(x)g(z)}[x,[y,z]] + (-1)^{g(y)g(x)}[y,[z,x]]
 + (-1)^{g(z)g(y)}[z,[x,y]] =0
\eqn\zb
$$
for any $x$, $y$ and $z$ in ${\cal G}$,
where $g(x)=0$ if $x\in {\cal G}_B$
and $g(x)=1$ if $x\in {\cal G}_F$.
 ${\cal G}$ will be  assumed to possess
a real, non-degenerate, supersymmetric
bilinear form $(\,,)$  such that ${\cal G}_B$
and ${\cal G}_F$  are  orthogonal. It will also be
assumed to satisfy the {\it invariance
property}
$$
([x,y],z)=(x,[y,z])
\eqn\zc
$$
for all $x$, $y$ and $z$ in ${\cal G}$. Call
$g_{ab}=(T_a,T_b)$, $ g_{\alpha\beta}=(T_{\alpha},T_{\beta})$;
one has
$g_{ab}=g_{ba}$, $g_{\a\b}=-g_{\b\a}$.
Indices are raised and
lowered by contraction with the metric
tensor and its inverse according to
$T^a=g^{ab}T_b$, $T_a=g_{ab}T^b$, $T^{\a}=T_{\b}g^{\b\a}$,
$T_{\a}=T^{\b}g_{\b\a}$. Upon lowering of its upper index,
$$
f_{abc}=f_{ab}^{d}g_{dc},\quad
f_{a\b\g}=f_{a\b}^{\mu}g_{\mu\g},\quad
f_{\a\b c}=f_{\a\b}^{d}g_{dc},
\eqn\zd
$$
the structure constants become superantisymmetric. The Jacobi identity
(eq. \zb) imposes additional conditions on the structure constants.
Some of these conditions are:
$$
\eqalign{
&f_{\alpha\beta}^{c}\,f_{ca}^{d}\,-\,
f_{\beta a}^{\lambda}\,f_{\lambda \alpha}^{d}\,+\,
f_{ a\alpha}^{\lambda}\,f_{\lambda \beta}^{d}\,=\,0\cr
&f_{ab}^{c}\,f_{c\alpha}^{\beta}\,+\,
f_{b\alpha}^{\lambda}\,f_{\lambda a}^{\beta}\,+\,
f_{ \alpha a}^{\lambda}\,f_{\lambda b}^{\beta}\,=\,0\cr
&f_{\alpha\beta}^{c}\,f_{c\gamma}^{\delta}\,+\,
f_{\beta \gamma}^{c}\,f_{c \alpha}^{\delta}\,+\,
f_{ \gamma\alpha}^{c}\,f_{c \beta}^{\delta}\,=\,0\cr}
\eqn\zh
$$
Other relations satisfied by the structure constants can be
obtained from the quadratic Casimir operator of the
superalgebra. If $C_A$ denotes the value of this operator in
the adjoint representation, one has:
$$
\eqalign{
&g^{bc} f_{ab}^{d} f_{dc}^{e}\,+\,
g^{\alpha\beta} f_{a\alpha}^{\gamma}
f_{\gamma\beta}^{e}
=C_A\,\delta ^{e}_{a}\cr\cr
&g^{bc} f_{\alpha b}^{\gamma} f_{\gamma c}^{\mu}\, =\,
g^{\beta\gamma} f_{\alpha\beta}^{c} f_{c\gamma}^{\mu}\,=\,
{C_A\over 2}\,\, \delta ^{\mu}_{\alpha}\cr}
\eqn\zi
$$

A conformal current superalgebra is generated
by a set of holomorphic
bosonic  $J_a(z)$ and  fermionic currents $J_{\a}(z)$,
satisfying the following operator product expansions (OPE's):
$$
\eqalign{
J_a(z)J_b(w)=&{k g_{ab}\over (z-w)^2} +
{f_{ab}^{c}\over z-w}J_c(w)\cr
J_a(z)J_{\b}(w)=&{f_{a\b}^{\g}\over z-w}J_{\g}(w)\cr
J_{\a}(z)J_{\b}(w)=&{k g_{\a\b}\over (z-w)^2} +
{f_{\a\b}^{c}\over z-w}J_c(w).
\cr}
\eqn\ze
$$
$k$ is the level of the current superalgebra.
The currents $J_a$ and $J_{\a}$ can
be made into Virasoro primary  fields by
application of the  Sugawara
construction, whereby a bilinear in the $J$'s
$$
T^J=N(g^{ab}J_{a}J_{b} + g^{\a\b}J_{\a}J_{\b})
\eqn\zf
$$
is required to satisfy a Virasoro algebra,
such that all the currents $J$ have
conformal dimension 1 with respect to $T^J$;
the normalisation constant $N$ is fixed precisely by this
requirement. This analysis
is  standard and has been carried out,
for the case of a  Lie superalgebra,
\REF\goddard{P. Goddard, D. Olive and
G. Waterson\journal\cmp&112(87)591.}
\REF\jarvis{P. D. Jarvis and R. B. Zhang\journal\pl&B215(88)695.}
\REF\henningson{M. Henningson\journal\ijmp&A6(91)1137.}
\REF\troost{A. Deckmyn and W. Troost\journal\np&B370(92)231.}
\REF\fujitsu{A. Fujitsu\journal\mpl&A8(93)1763.}
in [\goddard,\jarvis,\henningson,\troost,\fujitsu]. The
value of $N$ is found to be
$$
N={1\over 2k + C_A},
\eqn\zg
$$
and the operator
$$
T^J={1\over 2k + C_A}(g^{ab}J_{a}J_{b}
+ g^{\a\b}J_{\a}J_{\b})
\eqn\zj
$$
closes a Virasoro algebra with a central charge $c_J$ given
by $$
c_J= {2k (d_B - d_F)\over 2k+ C_A}.
\eqn\zk
$$
Let us now describe how one can construct a TCFT based on
the algebra \ze. The basic ingredient in our construction
is the introduction of a ghost sector such that one can
realise the BRST symmetry of the superalgebra \ze. We
shall use this BRST symmetry as the topological symmetry
of the TCFT. In general, in order to represent the BRST
symmetry of a given chiral algebra, one has to introduce
a fermionic(bosonic) ghost system for every
bosonic(fermionic) generator of the algebra. To these
ghost systems one must assign ghost numbers in such a way
that each antighost and its associated generator have the
same spin. According to these general rules we must
introduce in our case a spin-one ghost system for each
current of the superalgebra \ze. Moreover, the
topological symmetry we are trying to implement is such
that the BRST variation of the antighost equals the
corresponding total current.

Let us denote by $(\rho_a,\gamma^a)$ to the fermionic
ghosts for the currents $J_a$ whereas the bosonic fields
$(\lambda_{\alpha}, \eta^{\alpha})$ will correspond to
the currents $J_{\alpha}$. Let us choose our conventions
in such a way that the fermionic ghost fields $\rho _{a}$,
$\gamma ^{b}$ satisfy the OPE
$$
\rho_a(z) \gamma^b(w)= {-\delta^{b}_{a}\over z-w}.
\eqn\zl
$$
$\rho_a$ and $\gamma^b$ will be assumed to have
 conformal weights 1 and 0,
and will be assigned ghost numbers $-1$ and $+1$,
respectively. Their
energy-momentum tensor $T^{(\rho\gamma)}$ is given by
$$
T^{(\rho\gamma)}=\rho_{a}\partial \gamma^{a},
\eqn\zm
$$
and has a central charge $c_{(\r\g)}=-2d_B$.
Similarly the bosonic ghost fields $\eta ^{\a}$,
$\lambda _{\b}$ will satisfy
$$
\eta^{\a}(z) \l_{\b}(w)= {-\delta^{\a}_{\b}\over z-w}.
\eqn\zn
$$
$\eta ^{\a}$ and $\lambda _{\b}$ have  conformal weights 0
and 1 and ghost numbers $+1$ and $-1$, respectively,
and  their energy-momentum tensor is:
$$
T^{(\e\l)}=\partial \eta^{\a} \l_{\a}.
\eqn\zo
$$
The central charge of   $T^{(\e\l)}$  is
$c_{(\eta\l)}=2d_F$. Altogether,  the operator
$$
T=T^J + T^{(\r\g)} + T^{(\eta\l)}
\eqn\zp
$$
closes a Virasoro algebra with a central charge given by
$$
c_{(tot)}=[\,{2k \over 2k+ C_A} -
2]\,(d_B - d_F).
\eqn\zq
$$

A TCFT is expected to have a vanishing Virasoro anomaly.
Notice that the condition $c_{(tot)}=0$ is satisfied when
$k=-C_A$. This value of the level determines the
topological point of the current + ghost system under
consideration. Another way to see how a
 topological theory comes about is the following. Let us
consider the  combination
$$
J_{a}^{{\rm gh}}=f_{ab}^{c}\g^{b}\r_{c} -
f_{a\b}^{\g}\eta^{\b}\l_{\g},
\eqn\zr
$$
which is a bosonic, zero ghost-number field with
conformal weight 1 with
respect to $T$ in eq. \zp. Similarly, consider the object
$$
J_{\a}^{{\rm gh}}= f_{\a\mu}^{b}\r_{b}\eta^{\mu} +
f_{\a b}^{\mu}\g^{b}\l_{\mu},
\eqn\zs
$$
with the same quantum numbers as above,
but fermionic statistics.
$J_{a}^{{\rm gh}}$ and $J_{\a}^{{\rm gh}}$
can be checked to verify the
algebra \ze, with a level  $k=C_A$. Actually one can
easily verify that $J_{a}^{{\rm gh}}$ and
$J_{\a}^{{\rm gh}}$ represent the generators of the
superalgebra in the space of ghost fields. Therefore
 the {\it total} bosonic and
fermionic currents  read
$$
{\cal J}_{a}= J_{a}+J_{a}^{{\rm gh}}\qquad
{\cal J}_{\a}=J_{\a}+J_{\a}^{{\rm gh}},
\eqn\zt
$$
and satisfy the algebra \ze\ with a
total level $k_{(tot)}= k + C_A$.
According to the general arguments of refs. [\tca,
\tkt], a topological  current superalgebra should now
appear at that particular value of $k$ for which
$k_{(tot)}$ vanishes, \ie, for
$k=-C_A$. That this is indeed
correct is confirmed by the fact that
this latter value, when substituted into
eq. \zq, gives $c_{(tot)}=0$.

We now work out the topological structure
present in the theory. To begin
with, a nilpotent BRST current $Q$ having
fermionic statistics, conformal
weight 1 and ghost number $+1$ is needed.
With the fields at hand, the
combination
$$
Q=-\g^a J_a - {1\over 2}f_{ab}^{c}\g^a\g^b\r_c +
f_{a\b}^{\mu}\g^a\eta^{\b}\l_{\mu} - \eta^{\a} J_{\a}
 - {1\over 2}f_{\a\b}^{c}
\eta^{\a}\eta^{\b}\r_c
\eqn\zu
$$
satisfies the necessary requirements. Indeed $Q$ is the
canonical BRST charge for the Lie superalgebra \ze.
A tedious
although straightforward
calculation shows that
$$
Q(z)Q(w) = {k+C_A\over z-w}\,(g_{ab}\partial\g^a\g^b +
g_{\a\b}\partial \eta^{\a}\eta^{\b}),
\eqn\zv
$$
thus confirming again that only
for the critical value $k=-C_A$ is it
possible to have a nilpotent topological symmetry.
 From now on we will always
assume that we are working at the critical level.

One can now suspect that all the operators present
in the theory appear in
BRST doublets. To prove that this statement is true,
let us begin with the
total Ka\v c--Moody currents as given in eq. \zt. We have
$$
\eqalign{
Q(z)\r_a(w)=&{1\over z-w}{\cal J}_a(w)\cr
Q(z){\cal J}_a(w)=&0\cr
Q(z)\l_{\a}(w)=&{1\over z-w}{\cal J}_{\a}(w)\cr
Q(z){\cal J}_{\a}(w)=&0,\cr}
\eqn\zw
$$
so $(\r_a,{\cal J}_a)$ and $(\l_{\a},{\cal J}_{\a})$
form weight 1 topological doublets. In eq. \zw\ one
notices that the BRST variations of the antighosts
$\r_a$ and $\l_{\a}$ are the total currents ${\cal J}_a$
and ${\cal J}_{\a}$ respectively, which confirms the
correctness of our choice for $Q$. The operator algebra
of $\r_a$, $\l_{\a}$, ${\cal J}_b$ and
${\cal J}_{\beta}$ closes as follows:
$$
\eqalign{
{\cal J}_a(z){\cal J}_b(w)=
&{f_{ab}^c\over z-w}{\cal J}_c(w)\cr
{\cal J}_a(z){\cal J}_{\b}(w)=
&{f_{a\b}^{\g}\over z-w}{\cal J}_{\g}(w)\cr
{\cal J}_{\a}(z){\cal J}_{\b}(w)=
&{f_{\a\b}^{c}\over z-w}{\cal J}_c(w)\cr
{\cal J}_a(z)\r_b(w)=&{f_{ab}^c\over z-w}\r_c(w)\cr
{\cal J}_a(z)\l_{\a}(w)=&{f_{a\a}^{\b}\over z-w}\l_{\b}(w)\cr
{\cal J}_{\a}(z)\r_b(w)=&-{f_{\a b}^{\g}\over z-w}\l_{\g}(w)\cr
{\cal J}_{\a}(z)\l_{\b}(w)=&-{f_{\a\b}^c\over z-w}\r_c(w).\cr}
\eqn\zx
$$
The topological character of the theory is ensured if the
energy-momentum tensor $T$ in eq. \zp\ is $Q$-exact. In
that case the BRST ancestor of $T$, denoted by $G$,
would  be a weight 2 fermionic field with
ghost number $-1$. A glance at
eq. \zw\ can give some idea about
its expression: a Sugawara-like bilinear
of the form $g^{ab}\r_a J_b$, $g^{\a\b}\l_{\a}J_{\b}$, with some
appropriate coefficients, will do. The precise combination
$$
G= {-1\over C_A} (g^{ab}\r_a J_b + g^{\a\b}\l_{\a}J_{\b}),
\eqn\zz
$$
where the overall coefficient equals the one
in eq. \zj\ for the
critical level, satisfies
$$
Q(z)G(w)= {d_B-d_F\over (z-w)^3} +
{1\over (z-w)^2}R(w) + {1\over z-w}T(w).
\eqn\zaa
$$
In eq. \zaa, $R$ is a weight 1 bosonic
field with ghost number zero given by
$$
R=\r_a\g^a + \l_{\a}\eta ^{\a}.
\eqn\zab
$$
Eq. \zaa\ is characteristic of two-dimensional
topological models. The
residue at the simple pole is the energy-momentum
tensor, which proves its
BRST-exactness, while the operator appearing
at the double pole is a
$U(1)$ current. Indeed, one easily checks that
$$
R(z)R(w)= {d_B-d_F\over (z-w)^2},
\eqn\zac
$$
and $R$ can be understood as the ghost-number
 current. Indeed the $R$-charges of the different fields
coincide with the ghost numbers we have assigned them.
As for the triple pole
in eq. \zaa, the coefficient is a c-number
called the {\it topological
dimension} $d$ of the model, which now equals
the superdimension of the current
superalgebra. The same arguments as those
developed in [\tca] lead us to
conclude
that we are in fact describing a topological
sigma model for the underlying
supergroup manifold. A second BRST-doublet is
thus given by $(G,T)$, and the
OPE
$$
Q(z)R(w)={-1\over z-w}Q(w)
\eqn\zad
$$
shows that $(R,Q)$ too are BRST partners.
One can now compute the remaining
algebra between  the generators above, with the result that
$$
\eqalign{
T(z)Q(w)=&{1\over (z-w)^2}Q(w) + {1\over z-w}\partial Q(w)\cr
T(z)R(w)=&{-(d_B-d_F)\over (z-w)^3} + {1\over (z-w)^2}R(w) +
 {1\over z-w}
\partial R(w)\cr
T(z)G(w)=&{2\over (z-w)^2}G(w) + {1\over z-w}\partial G(w)\cr
R(z)G(w)=&{-1\over z-w}G(w).\cr}
\eqn\zae
$$
The above OPE's are exactly
those obtained upon twisting the $N=2$
superconformal algebra
\REF\LVW{W. Lerche, C. Vafa and N. P. Warner
\journal\np&B324(89)427.}
[\LVW,\EY] (the so-called {\it topological algebra}),
so one might be led to believe that such an
algebra is also present here. However, there is a
fundamental difference now, because the BRST partner of
$T$, $G$, is {\it not}
nilpotent. Instead one has
$$
G(z)G(w)={1\over z-w}W(w),
\eqn\zaf
$$
where $W$ is a bosonic, dimension 3 operator
with ghost number $-2$ given by
$$
\eqalign{
W=&-{1\over C_A}(\partial \r_a\r^a +
\partial \l_{\a}\l^{\a})\cr
 +&{1\over (C_A)^2}\,(f_{ab}^c\r^a \r^b J_c +
f_{\a \b}^c \l^{\a} \l^{\b} J_c +
2 f_{a\b}^{\g} \r^a \l^{\b} J_{\g} ).\cr}
\eqn\zag
$$
Since all operators in the theory so far
have appeared as BRST doublets, one
would expect this to hold for $W$, too.
And given that $W$ is BRST-closed, \ie,
$$
Q(z)W(w)=0,
\eqn\zah
$$
we must look for a BRST ancestor for $W$
with the following quantum numbers:
fermionic
statistics, conformal weight 3, and ghost
number $-3$. The combinations
$f_{abc}\r^a \r^b \r^c$ and $f_{\a\b c}\l^{\a} \l^{\b} \r^c$
immediately
come to mind. Defining
$$
V={1\over (C_A)^2}\,\,\Bigl({1\over 3}f_{abc}\r^a \r^b \r^c +
f_{\a\b c}\l^{\a} \l^{\b} \r^c\Bigr),
\eqn\zai
$$
one can check that
$$
Q(z)V(w)={1\over z-w}W(w),
\eqn\zaj
$$
which proves our point: $(V,W)$ forms a new BRST doublet.

In trying to work out the topological
 structure present in the theory, we have
found that the operator algebra is very similar
to that of the twisted $N=2$ models.
But the appearance of $(V,W)$ forces us to compute
their OPE's with all other
operators, and there is no guarantee that
the resulting algebra will close on
a finite number of fields. However, the
algebra of $(G,T)$, $(R,Q)$ and
$(V,W)$ does close, as some computation proves. The results are
$$
\eqalign{
T(z)W(w)=&{3\over (z-w)^2}W(w) + {1\over z-w}\partial W(w)\cr
T(z)V(w)=&{3\over (z-w)^2}V(w) + {1\over z-w}\partial V(w)\cr
G(z)W(w)=&{3\over (z-w)^2}V(w) + {1\over z-w}\partial V(w)\cr
R(z)W(w)=&{-2\over z-w}W(w)\cr
R(z)V(w)=&{-3\over z-w}V(w),\cr}
\eqn\zak
$$
while all other OPE's vanish identically.
It is important to emphasise that,
contrary to what happens with ${\cal W}$ algebras,
the existence of
higher-spin fields does not spoil the linearity of the algebra.

The above conclusions have also been obtained in
[\tca,\gln], but our
analysis here extends these results to the
more general case of an arbitrary
superalgebra. It should also be mentioned
that the algebra exhibited in
eqs. \zaa\ to \zak, which we shall call the Kazama
algebra, first appeared in [\kazama] as a consistent,
non-trivial  extension of the twisted $N=2$ algebra.
In [\kazama] it was related to an $N=1$
superconformal symmetry, but no explicit
representation for the generators was given
(see also ref.
\REF\getzler{E. Getzler
\journal\ap&237(95)161.}[\getzler]).

To complete our analysis, it remains to
study whether or not the currents
${\cal J}_a$, ${\cal J}_{\a}$ and their
BRST ancestors $\r_a$, $\l_{\a}$,
on the one hand, and the generators $T$,
$G$, $R$, $Q$, $W$ and $V$,
on the other, are compatible. Some of the
corresponding OPE's are trivial
(for example, those expressing the Virasoro primary
character of the currents); others have already
been given (eq. \zw).
Among those remaining, the only non-vanishing ones are
$$
\eqalign{
G(z){\cal J}_a(w)=&{1\over (z-w)^2}\r_a(w) +
{1\over z-w}\partial \r_a(w)\cr
G(z){\cal J}_{\a}(w)=&{1\over (z-w)^2}\l_{\a}(w) +
{1\over z-w}\partial \l_{\a}(w)\cr
R(z)\r_a(w)=&{-1\over z-w} \r_a(w)\cr
R(z)\l_{\a}(w)=&{-1\over z-w}\l_{\a}(w),\cr}
\eqn\zal
$$
which establishes that the  topological and
current superalgebra structures are
indeed compatible.

An interesting feature of the above
construction is the fact that it can also
be performed with two independent sets of
currents $J^{1}$, $J^{2}$. Suppose
$J^{1}$ and $J^{2}$ satisfy the algebra \ze\
with levels $k_1$ and $k_2$,
respectively. Then the Sugawara energy-momentum
tensor $T^J$ is given by
$$
T^J=
{1\over 2k_1 + C_A}(g^{ab}J^{1}_{a}J^{1}_{b} +
g^{\a\b}J^{1}_{\a}J^{1}_{\b}) +
{1\over 2k_2 + C_A}(g^{ab}J^{2}_{a}J^{2}_{b} +
g^{\a\b}J^{2}_{\a}J^{2}_{\b}),
\eqn\zam
$$
and the corresponding central charge is
$$
c_J= [\,{2k_1 \over 2k_1+ C_A}\,
          +\, {2k_2 \over 2k_2+ C_A }]\,(d_B - d_F).
\eqn\zan
$$
Imposing $k_1+k_2=-C_A$ and
setting $k_1=k$ for simplicity,
eqs. \zam\ and \zan\ reduce to
$$
T^J=
{1\over 2k + C_A}\bigr[g^{ab}(J^{1}_{a}J^{1}_{b}-
J^{2}_{a}J^{2}_{b})+
g^{\a\b}(J^{1}_{\a}J^{1}_{\b} -J^{2}_{\a}J^{2}_{\b})\bigl]
\eqn\zao
$$
and
$$
c_J=2(d_B-d_F),
\eqn\zap
$$
which exactly cancels the ghost central
charge $c_{(\r\g)}+c_{(\eta\l)}$. That
this is indeed a new topological point
is again confirmed by the following
arguments. The new total currents are
$$
\eqalign{
{\cal J}_a=&J^1_a + J^2_a + J^{{\rm gh}}_a\cr
{\cal J}_{\a}=&J^1_{\a} + J^2_{\a} + J^{{\rm gh}}_{\a},\cr}
\eqn\zaq
$$
with $J^{{\rm gh}}$ as in eqs. \zr, \zs, and their algebra
is  $$
\eqalign{
{\cal J}_a(z){\cal J}_b(w)=&{(k_1+k_2+C_A) g_{ab}
\over (z-w)^2} +
{f_{ab}^{c}\over z-w}{\cal J}_c(w)\cr
{\cal J}_a(z){\cal J}_{\b}=&{f_{a\b}^{\g}\over z-w}
{\cal J}_{\g}(w)\cr
{\cal J}_{\a}(z){\cal J}_{\b}(w)=&{(k_1+k_2+C_A)
g_{\a\b}\over (z-w)^2} +
{f_{\a\b}^{c}\over z-w}{\cal J}_c(w).\cr}
\eqn\zar
$$
The new BRST current making them BRST-exact is
$$
Q=-\g^a (J^1_a+J^2_a) - {1\over 2}f_{ab}^{c}\g^a\g^b\r_c +
f_{a\b}^{\mu}\g^a\eta^{\b}\l_{\mu} - \eta^{\a} (J^1_{\a}+J^2_{\a}) -
{1\over 2}f_{\a\b}^{c}\eta^{\a}\eta^{\b}\r_c,
\eqn\zas
$$
where again $\r_a$ and $\l_{\a}$ are their BRST ancestors.
 Nilpotency of $Q$
occurs only at the topological point, since
$$
Q(z)Q(w) = {k_1+k_2+C_A\over z-w}\,(g_{ab}\partial\g^a\g^b +
g_{\a\b}\partial \eta^{\a}\eta^{\b}).
\eqn\zat
$$
One can now repeat the above analysis and work out the expressions for all the
operators, with the result that the
 topological algebra is satisfied without
changes. The ghost number current $R$ and the
 topological dimension $d$ remain the same, but the
other generators have to be modified as follows:
$$
\eqalign{
G=& {1\over 2k+C_A}\,\, \Bigl[g^{ab}\r_a (J^1_b-J^2_b) +
g^{\a\b}\l_{\a}(J^1_{\b}-J^2_{\b})\Bigr]\cr
W=&{1\over (2k+C_A)^2}\,\,\Bigl[-C_A\,(\partial \r_a\r^a +
\partial \l_{\a}\l^{\a}) +
2 f_{a\b}^{\g} \r^a \l^{\b} (J^1_{\g}+J^2_{\g})\cr
+&f_{ab}^c\r^a \r^b (J^1_c+J^2_c) +
f_{\a \b}^c \l^{\a} \l^{\b} (J^1_c+J^2_c)\Bigr]
\cr
V=&{1\over (2k+C_A)^2}\,\,
\Bigl[{1\over 3}f_{abc}\r^a \r^b \r^c +
f_{\a\b c}\l^{\a} \l^{\b} \r^c\Bigr].\cr}
\eqn\zau
$$
Although the topological algebra is
the same as in the one-current case, this
two-current construction is interesting
 because it allows for a lagrangian
interpretation of the theory as a
${\cal G}/{\cal G}$ coset. This point  is
examined in the next section.
Before finishing this one, let us point out that the topological
algebra we  have studied admits deformations both in its one and two
current realizations. Indeed, if $\alpha^a$ are c-number constants, one
can redefine  $T$, $G$ and $R$ as follows:
$$
\eqalign{
T&\rightarrow T+\sum_a\alpha^a\partial {\cal J}_a\cr
G&\rightarrow G+\sum_a\alpha^a\partial \rho_a\cr
R&\rightarrow R+\sum_a\alpha^a {\cal J}_a,\cr}
\eqn\zauu
$$
The operators $Q$, $V$ and $W$ are left unaffected by the deformation.
One easily checks that the transformed generators  satisfy the extended
topological algebra for any value of the $\alpha^a$ constants. Of
course, after the deformation, the currents are no longer primary
dimension-one operators. Transformations of the type displayed in
eq.\zauu\ will play an important role in section 4, where we shall
relate our results with the non-critical string theories.

\chapter{A gauged, supergroup-valued WZW model.}

The topological algebra described in
the previous section has a lagrangian interpretation that
we now develop. We shall show below that it is possible to give a
lagrangian description of the  two-current
construction of section 2. The main
result of this section is the
interpretation of the gauged, ${\cal G}$-valued WZW model as a
theory in which the extended topological algebra closed by
$(T,G)$, $(Q,R)$ and $(W,V)$ is realised. A similar conclusion has
also been reported in [\tca] for the bosonic case (\ie, when
$d_F=0$), but our presentation here is totally general. An earlier
reference on gauged WZW models is \REF\schnitzer{H. J.
Schnitzer\journal\np&B324(89)412.} [\schnitzer]. For arbitrary
supergroups, a lagrangian construction of ${\cal G}/{\cal G}$ has
already been put forward \REF\yu{J. B. Fan and M. Yu, ``G/G Gauged
Supergroup Valued WZNW Field Theory", Academia Sinica preprint
AS-ITP-93-22.} in [\yu].

Our starting point is the ${\cal G}$-gauged WZW functional
$$
\Gamma (g,A) = \,\Gamma (g) -
{1\over \pi}\int_{\Sigma}d^2z\,
{\rm str}(g^{-1}A_{\bar z}gA_z - A_{\bar z}\partial_zg g^{-1} +
g^{-1}\partial_{\bar z}gA_z - A_zA_{\bar z}),
\eqn\zav
$$
with $\Gamma (g)$ given by
$$
\Gamma (g) = {1\over 2 \pi}\int_{\Sigma}d^2z\,{\rm
str}(g^{-1}\partial_zgg^{-1}\partial_{\bar z}g) + {i\over
12\pi}\int_{M}\epsilon^{\mu\nu\rho}\,{\rm
str}(g^{-1}\partial_{\mu}g
g^{-1}\partial_{\nu}gg^{-1}\partial_{\rho}g).
\eqn\zaw
$$
$g$ is a function taking values in
the supergroup whose Lie superalgebra is ${\cal G}$, and the
3-manifold $M$ is such that $\partial M = \Sigma$.
The {\it na\"{\i}ve} partition function is
$$
Z = \int Dg DA_{\bar z}DA_z \exp{[-k \Gamma (g,A)]}.
\eqn\zax
$$
The gauge invariance of $\Gamma (g,A)$ is well known.
Also useful is the Polyakov-Wiegmann identity
satisfied by $\Gamma (g)$,
$$
\Gamma (gh) = \Gamma (g) + \Gamma (h) + \langle g,h\rangle ,
\eqn\zay
$$
where
$$
\langle g,h\rangle = {1\over \pi}\int_{\Sigma}d^2z\,
{\rm str}(g^{-1}\partial_{\bar z}g\,\partial_z
hh^{-1}).
\eqn\zaz
$$
Parametrise the gauge fields as
$$
A_{\bar z}=h^{-1}\partial_{\bar z}h\qquad
 A_z={\bar h}^{-1}\partial_z \bar h
\eqn\zba
$$
with $h$ and $\bar h$ taking values in the supergroup,
and change variables in the functional
integral \zax. One has
$$
DA_zDA_{\bar z}=\,J[h,\bar h]\,DhD\bar h,
\eqn\zbb
$$
where $J[h,\bar h]$ is the Jacobian for the change of
variables $A_z,\,\, A_{\bar z}\rightarrow h,\,\, \bar h$.
This Jacobian can be represented as a functional integral
over ghost fields that take values in the adjoint
representation of  ${\cal G}$. Denoting, as in the
previous section, these ghost fields by $(\r_a,\g^a)$
and $(\l_{\alpha},\eta^{\alpha})$,
 we have:
$$
J[h,\bar h]\,=\,\exp {[C_A
\Gamma (h^{-1}\bar h)]}\,
\int D\r D\g D\l D\eta \,
\exp{[{-1\over \pi}\int_{\Sigma} d^2z\,
(\r_a\partial_{\bar z}\g^a +
\l_{\alpha}\partial_{\bar z}\eta ^{\alpha} + {\rm
c.c.})]}.
\eqn\extra
$$
Taking into account that
$$
\Gamma (g,A) = \,\Gamma (h^{-1}g\bar h) -
\Gamma (h^{-1}\bar h),
\eqn\extrados
$$
the partition function becomes
$$
\eqalign{
Z=\int DgDhD\bar h  D\r D\g D\l D\eta
\,&\exp{[-k \Gamma (h^{-1}g\bar h) +
 (k+ C_A)\Gamma(h^{-1}\bar h)]}\cr
&\exp{[{-1\over \pi}\int_{\Sigma} d^2z\,
(\r_a\partial_{\bar z}\g^a +
\l_{\alpha}\partial_{\bar z}\eta ^{\alpha} +
{\rm c.c.})]}.\cr}
\eqn\zbc
$$
Changing variables as $h^{-1}g\bar h\rightarrow g$ and
choosing the gauge $h=1$ (which does not introduce any
new Faddeev-Popov ghosts) we obtain:
$$
\eqalign{
Z=\int DgD\bar h D\r D\g D\l D\eta
&\exp{[-k \Gamma (g) + (k+C_A)\Gamma(\bar h)]}\cr
&\exp{[{-1\over \pi}\int_{\Sigma} d^2z\,
(\r_a\partial_{\bar z}\g^a +
\l_{\alpha}\partial_{\bar z}\eta ^{\alpha} +
{\rm c.c.})]}.\cr}
\eqn\zbd
$$
This gauge-fixed form for the partition function clearly
exhibits the necessary elements to
construct the ${\cal G}/{\cal G}$ theory,
as realised with two currents: one needs two independent,
ungauged WZW models such that their levels add up to $-C_A$, plus a
compensating ghost sector in order to set the total level to zero. Let
us finally point out that the lagrangian interpretation we have
discussed in this section allows to interpret the BRST symmetry of the
superalgebra ${\cal G}$ as the basic symmetry of a topological sigma
model having a Lie supergroup as target space.

\chapter{Relation with non-critical superstrings.}

In this section we shall explore the relation between the
topological theories constructed in the previous sections and the
non-critical superstring theories. In particular, we will argue
that, for the Lie superalgebras ${\rm osp}(1|2)$ and ${\rm osp}(2|2)$, the
corresponding topological coset models are related to the $N=1$
and $N=2$ superstring theories, respectively. In order to find
this correspondence one must first conveniently deform the
two-current model, as was explained at the end of section 2.
In the deformed theory one can implement a quantum
Drinfeld-Sokolov hamiltonian reduction in such a way that the
reduced model can be identified with the corresponding string
theory.

The connection of the topological current system with non-critical
strings is more transparent if a free field realisation of the
two currents involved in the ${\cal G}/ {\cal G}$ coset is
used. Roughly speaking, one can associate one of the two currents
with the matter sector of the string whereas the other current is
related to the Liouville degrees of freedom. Moreover,
 some of the ghosts of the deformed ${\cal G}/ {\cal
G}$ coset can be identified with those of string theory. The
remaining fields coming from the ghost and current sectors can be
organised into topological quartets and one can invoke the
standard Kugo-Ojima confinement mechanism to eliminate
these quartets from the physical Hilbert space.

Before studying the ${\rm osp}(1|2)$ and ${\rm osp}(2|2)$ cases, let us
for completeness
recall
[\yank,\hu]  the relation between the ${\rm sl}(N)/{\rm  sl}(N)$
cosets and the non-critical $W_N$-strings. First of all let us
briefly describe the root system of the ${\rm sl}(N)$ Lie algebra. If
$\vec e_i$ is a unitary vector in ${\RR}^N$ along the ${\rm
i}^{{\rm th}}$ axis and $\vec \epsilon_{ij}\,=\,\vec e_i\,-\,\vec
e_j$, then the positive roots of ${\rm sl}(N)$ are the elements of the set
$\Delta_+\,=\,\{\,\vec \epsilon_{ij},\,\,\,j>i\,\}$, while the
simple roots are given by
$\vec \alpha_i\,=\,\vec\epsilon_{i,i+1}$ ($i=1,\cdots,N-1)$. Any
positive root $\vec \alpha\in\Delta_+$ can be written as
$\vec \alpha\,=\,\sum_{i=1}^{N-1}\,n_{\alpha}^i\vec\alpha_i$
where the $n_{\alpha}^i$'s are non-negative integers. The height
of $\vec \alpha$ is given by:
$$
h_{\alpha}\,=\,\sum_{i=1}^{N-1}\,n_{\alpha}^i.
\eqn\uno
$$
In particular, for $j>i$, $h_{\epsilon_{ij}}\,=\,j-i$. Denoting
 by $\vec \delta$  half the sum of positive roots
(\ie\
$\vec \delta\,=\,{1\over 2}\,\sum_{\alpha\in\Delta_+}\,
\vec \alpha$),  the height of any $\vec \alpha\in\Delta_+$ is
simply given by
$h_{\alpha}\,=\,\vec\delta\cdot\vec\alpha$.

We adopt the free field realisation of the ${\rm sl}(N)$ current
algebra of ref.
\REF\gera{A. Gerasimov et al. \journal\ijmp&A5(90)2495.}
[\gera]. In this realisation, a spin-one
bosonic $\beta\gamma$ system is introduced for every positive
root $\vec \alpha\in\Delta_+$. If we call
these bosonic fields
 $(w_{\alpha},
\chi_{\alpha})$, then the expression for the
currents associated with the positive roots is of the form:
$$
J_{\alpha}\,=\,w_{\alpha}+\cdots,
\eqn\dos
$$
where the dots denote terms which are non-linear in the fields.
Notice that the conformal weights of $(w_{\alpha},
\chi_{\alpha})$ are $\Delta (w_{\alpha})=1$ and
$\Delta (\chi_{\alpha})=0$. One also needs to introduce a set of
$N-1$
scalar fields $\vec \phi\,=\,(\phi_1,\cdots, \phi_{N-1}\,)$.
 Then the expression for the Cartan currents can be
given in general. In fact if we represent by $\vec\mu\cdot\vec H$
the current along the direction of an arbitrary Cartan vector
$\vec\mu$, we have:
$$
\vec H\,=\,i\sqrt{k+N}\,\,\partial\vec\phi\,-\,
\sum_{\alpha\in\Delta_+}\,\vec\alpha\,w_{\alpha}
\chi_{\alpha},
\eqn\tres
$$
where $k$ is the level of the ${\rm sl}(N)$ algebra. It is also
possible to give a simple expression for the Sugawara
energy-momentum tensor in terms of these free fields:
$$
\eqalign{
T^J\,=&\,{1\over
2(k+N)}\,g^{ab}J_{a}J_{b}=\cr
=&\sum_{\alpha\in\Delta_+}\,w_{\alpha}\partial\chi_{\alpha}
-{1\over 2}\,(\partial \vec\phi)^2\,-\,
{i\over\sqrt{k+N}}
\,\vec\delta\cdot\partial^2\vec\phi.\cr}
\eqn\cuatro
$$
A simple calculation using the Freudenthal-de Vries ``strange"
formula ($12\, \vec \delta^2=N(N^2-1)$) shows that
the energy-momentum tensor in eq. \cuatro\ indeed has the correct
central charge for an affine algebra at level $k$.

The topological ${\rm sl}(N)/ {\rm sl}(N)$ coset is obtained by
combining two ${\rm sl}(N)$ currents with levels $k_1=k$ and
$k_2=-k-2N$. Let the corresponding free fields carry the labels
$1$ and $2$. We must also add a pair of fermionic ghosts for each
independent current direction. In what follows we shall denote
the ghost along the Cartan direction by $(\rho_i, \gamma^i)$
($i=1,\cdots,N-1$) and those associated with the positive
(negative) roots of the algebra by
$(\rho_{\alpha},\gamma^{\alpha})$
( $(\rho_{-\alpha},\gamma^{-\alpha})$ respectively).

In order to make contact with non-critical string theory we must
first deform the theory. It turns out that the appropriate
deformation of the total energy-momentum tensor $T$ is:
$$
T_{{\rm improved}}\,=\,T\,+\,\vec\delta\cdot\partial\vec
{\cal  H}.
\eqn\cinco
$$
In eq. \cinco\ $\vec {\cal  H}$ is the total Cartan current of the
${\rm sl}(N)/ {\rm sl}(N)$ coset (see eq. \zaq).
Let us separate in $T_{{\rm improved}}$ the contribution of the
currents from those of the ghosts:
$$
T_{{\rm improved}}\,=\,T_{{\rm improved}}^J+
T_{{\rm improved}}^{\rm gh}.
\eqn\seis
$$
Using the free-field representation of $T^J$ and
$\vec H$ given in eqs. \cuatro\ and \tres\ we can write the
explicit expression of $T_{{\rm improved}}^J$:
$$
\eqalign{
T_{{\rm improved}}^J\,=&\,
-{1\over 2}\,( \partial \vec\phi_1)^2\,
-{1\over 2}\,( \partial \vec\phi_2)^2\,+
i{t-1\over \sqrt{t}}\,\vec\delta\cdot\partial^2\vec\phi_1
\,-\,{t+1\over\sqrt{t}}\,
\vec\delta\cdot\partial^2\vec\phi_2\,+\cr
+&\,\sum_{i=1,2}\,\,\sum_{\alpha\in\Delta_+}[\,
(1-h_{\alpha})\,w_{\alpha}^i\partial\chi_{\alpha}^i
\,-\,h_{\alpha}\,\partial
w_{\alpha}^i\chi_{\alpha}^i\,],\cr}
\eqn\siete
$$
where $t=k+N$. Notice that, in the deformed theory, the fields
$(w_{\alpha},\chi_{\alpha})$  acquire a conformal weight that
depends on the height of the root $\vec \alpha$
($\Delta (w_{\alpha})=1-h_{\alpha}$,
$\Delta (\chi_{\alpha})=h_{\alpha}$). In order to compute the
ghost contribution to the improved energy-momentum tensor, we
need to know the part of $\vec {\cal H}$ that depends on the ghost
fields. From the commutation relations of ${\rm sl}(N)$ one easily gets:
$$
\vec {\cal H}^{\rm gh}\,=
\,\sum_{\alpha\in\Delta_+}\,\vec \alpha\,
[\,\gamma^{\alpha}\rho_{\alpha}-
\gamma^{-\alpha}\rho_{-\alpha}\,].
\eqn\ocho
$$
Using eq. \ocho\ one obtains
$$
\eqalign{
T_{{\rm improved}}^{\rm gh}=&\,\sum_{i=1}^{N-1}
\rho_i\partial \gamma^i\,+\,
\sum_{\alpha\in\Delta_+}[\,(1-h_{\alpha})\,\rho_{\alpha}
\partial\gamma^{\alpha}\,+\,h_{\alpha}\gamma^{\alpha}
\partial\rho_{\alpha}\,]\,+\cr
+&\sum_{\alpha\in\Delta_+}[\,(1+h_{\alpha})\,\rho_{-\alpha}
\partial\gamma^{-\alpha}\,-\,h_{\alpha}\gamma^{-\alpha}
\partial\rho_{-\alpha}\,].\,\cr}
\eqn\nueve
$$
The central charge of the field $\vec\phi_1$ can be computed from
the background charge displayed in eq. \siete. A simple
calculation shows that
$$
c_{\phi_1}\,=\,(N-1)\,[1\,-\,N(N+1)\,{(t-1)^2\over t}\,].
\eqn\diez
$$
When $t\,=\,{q\over p}$ with $q,p\in  \ZZ$ (\ie\ when
$k+N={q\over p}$), the central charge in eq. \diez\ is precisely
that of the minimal $(p,q)$ model of $W_N$ matter. It is also
easy to check from eq. \diez\ that  $\vec \phi_2$ has
the correct background charge to be considered as the
$W_N$-Liouville field. Moreover it can be seen that one can
always combine in a Kugo-Ojima topological quartet the $(w_{\alpha}^i,
\chi_{\alpha}^i)$ systems with ghost fields having the same conformal
weights. We can pair, for example, the $(w_{\alpha}^1, \chi_{\alpha}^1)$
fields with the ghosts $(\rho_{\alpha}, \gamma^{\alpha})$ corresponding
to the positive roots. Also the Cartan  ghosts
$(\rho_i, \gamma^i)$ can be paired with the fields
$(w_{\alpha}^2, \chi_{\alpha}^2)$ when  $\vec\alpha$ is a simple
root (\ie\ when $h_{\alpha}=1$), since in this case both systems
have dimensions $(1,0)$ and there are equal number of them. The
$(w_{\alpha}^2, \chi_{\alpha}^2)$ fields with $h_{\alpha}\geq 2$ can be
paired with some of the  $(\rho_{-\alpha}, \gamma^{-\alpha})$ ghosts.
By looking at the conformal weights of these last two systems one
concludes that, in order to combine $(w_{\alpha}^2, \chi_{\alpha}^2)$
with  $(\rho_{-\alpha'}, \gamma^{-\alpha'})$, the heights of $\vec
\alpha$ and $\vec\alpha'$ must satisfy
$h_{\alpha}-h_{\alpha'}=1$. A simple calculation tells one
 how many fields can be paired in this way. Since
the number of roots of ${\rm sl}(N)$ with height $h$ is $N-h$, the
difference between the number of
$(\rho_{-\alpha'}, \gamma^{-\alpha'})$ and
$(w_{\alpha}^2, \chi_{\alpha}^2)$ systems is
$N-h_{\alpha'}-(N-h_{\alpha})=h_{\alpha}-h_{\alpha'}=1$. It
follows that, for a given height $h$, there always remains one
unpaired $(\rho,\gamma)$ system with conformal weights
$(1+h,-h)$. We are thus left with a set of $N-1$
anticommuting ghost fields
with conformal weights $(2,-1),\cdots, (N,1-N)$.
Let us denote these fields by $(b_j,c_j)$ where the
conformal weight of $b_j$ is $j+1$ for $j=1,\cdots,N-1$. Notice
that they correspond to the ghost system of the $W_N$ string.
Therefore, writing $\vec \phi_M$ ($\vec \phi_L$) instead
of $\vec\phi_1$ ($\vec\phi_2$ respectively),
 the reduced energy-momentum tensor is given by:
$$
\eqalign{
T_{{\rm reduced}}\,=&\,
-{1\over 2}\,( \partial \vec\phi_M)^2\,
-{1\over 2}\,( \partial \vec\phi_L)^2\,+
i{t-1\over \sqrt{t}}\,\vec\delta\cdot\partial^2\vec\phi_M
\,-\,{t+1\over\sqrt{t}}\,
\vec\delta\cdot\partial^2\vec\phi_L\,+\cr
+&\,\sum_{j=1}^{N-1}\,\,[\,(j+1)b_j\partial c_j\,-\,jc_j\partial
b_j\,],\cr}
\eqn\once
$$
which, as  stated above, corresponds to that
 of matter coupled to $W_N$-gravity.

Next let us consider  the ${\rm osp}(1|2)$ Lie superalgebra. This
algebra contains three bosonic currents $J_{\pm}$ and $H$ that
close an ${\rm sl}(2)$ algebra at level $k$. In addition there are
 two fermionic currents that we shall denote by
$j_{\pm}$. The affine ${\rm osp}(1|2)$ superalgebra can be realised
[\bershadsky]
in terms of one scalar field $\phi$, one bosonic $\beta\gamma$ system
(denoted by  $(w,\chi)$), and one fermionic $bc$ system
 (denoted by $(\bar\psi, \psi)$), with dimensoins
 $\Delta(w)=\Delta (\bar\psi)=1$ and
$\Delta(\chi)=\Delta (\psi)=0$, satisfying the following basic OPE's:
$$
w(z)\,\chi(w)\,=\,\psi(z)\,\bar\psi(w)\,=\,{1\over z-w}
\,\,\,\,\,\,\,\,\,\,\,\,\,\,\,\,\,\,
\phi(z)\,\phi(w)\,=\,-{\rm log}\,(z-w).
\eqn\doce
$$
In terms of these fields the explicit form of the  ${\rm osp}(1|2)$
currents is:
$$
\eqalign{
J_+\,=&\,w\cr
J_-\,=&-\,w\chi^2\,+\,i\sqrt{2k+3}\,\,\chi\partial\phi\,-
\,\chi\psi\bar\psi\,+k\partial\chi\,+
\,(k+1)\psi\partial\psi\cr
H\,=&-w\chi\,+{i\over 2}\,\sqrt{2k+3}\,\,\partial\phi\,-\,
{1\over 2}\,\psi\bar\psi\cr
j_+\,=&\bar\psi\,+\,w\psi\cr
j_-\,=&-\chi(\bar\psi\,+\,w\psi)\,+i\sqrt{2k+3}\,\,
\psi\partial\phi\,+\,(2k+1)\partial\psi.\cr}
\eqn\trece
$$
Notice that for  ${\rm osp}(1|2)$, with our
conventions, $C_A=3$. Using this
value and the metric tensor extracted from OPE's of the currents in eq.
\trece, we can write the Sugawara energy-momentum tensor:
$$
T^J\,=\,{1\over
2k+3}\,[\,J_+J_-\,+\,J_-J_+\,+\, 2H^2\,-\,{1\over
2}\,j_+\,j_-\,+{1\over 2}\,j_-j_+\,].
\eqn\catorce
$$
Substituting the representation given in eq. \trece\ into
eq.\catorce\ one gets:
$$
T^J\,=\,w\partial\chi\,-\,\bar\psi\partial
\psi\,-\,{1\over 2}\,(\partial\phi)^2\,-\,
{i\over 2\sqrt{2k+3}}\,\partial^2\phi.
\eqn\quince
$$

In order to realise the topological ${\rm osp}(1|2)/{\rm osp}(1|2)$ coset
model we must combine two current systems with levels $k_1=k$
and $k_2=-k-3$. Let us denote the anticommuting ghost systems
for the currents $H$ and $J_{\pm}$ by $(\rho_0,\gamma^0)$ and
$(\rho_{\pm},\gamma^{\pm})$ respectively. The commuting ghosts
associated to the $j_{\pm}$ currents will be similarly denoted by
$(\lambda_{\pm},\eta^{\pm})$. In complete parallel with the
${\rm sl}(N)$ case, let us deform the energy-momentum tensor by adding
a derivative along the total Cartan current ${\cal H}$:
$$
T_{{\rm improved}}\,=\,T_{{\rm improved}}^{\rm J}\,+\,
T_{{\rm improved}}^{\rm gh}\,\equiv\,T\,
+\,\partial{\cal H}.
\eqn\dseis
$$
In eq. \dseis\ we have separated the contributions of the
currents from those of the ghosts. Let us consider
$T_{{\rm improved}}^{\rm J}$ first. Using eq. \quince\ one
immediately arrives at:
$$
\eqalign{
T_{{\rm improved}}^{\rm J}\,=&\,
-{1\over 2}\,(\partial\phi_1)^2\,
-\,{1\over 2}\,(\partial\phi_2)^2\,+\,
{i\over 2}\,{t-1\over\sqrt t}\,\partial^2\phi_1\,-\,
{1\over 2}\,{t+1\over\sqrt t}\,\partial^2\phi_2\,-\cr
-&\sum_{i=1,2}\,[\partial w^i\chi^i\,+\, {1\over 2}
\bar \psi^i\partial\psi^i\
+\, {1\over 2}
 \psi^i\partial\bar\psi^i\,],\cr}
\eqn\dsiete
$$
where now $t\,=\,2k+3$ and, as in the ${\rm sl}(N)$ case,
the indices $1$ and $2$ label the two currents. On the other
hand, from the basic OPE's of ${\rm osp}(1|2)$, the contribution of the
ghost fields to ${\cal H}$ is easily obtained. One gets:
$$
{\cal H}\,^{{\rm gh}}=
\,\gamma^+\rho_+\,-\,\gamma^-\rho_-\,- {1\over
2}\,\eta^+\lambda_+\,+\, {1\over 2}\,\eta^-\lambda_-.
\eqn\docho
$$
Taking eq. \docho\ into account it is straightforward to compute
$T_{{\rm improved}}^{\rm gh}\,$:
$$
\eqalign{
T_{{\rm improved}}^{{\rm gh}}\,=&
\,\rho_0\partial\gamma^0\,+\,
\gamma^+\partial\rho_+\,+\,2\rho_-\partial\gamma^- \,-\,
\gamma^-\partial\rho_-\cr
+&\,{1\over 2}\partial \eta^+\lambda_+
-\,{1\over 2} \eta^+\partial\lambda_+\,+\,
\,{3\over 2}\partial \eta^-\lambda_-\,+\,{1\over 2}
\eta^-\partial\lambda_-.\cr}
\eqn\dnueve
$$
Notice that in the deformed theory the conformal weights of the
antighost fields are:
$$
\Delta (\rho_0)\,=\,1
\,\,\,\,\,\,\,\,\,\,\,\,
\Delta (\rho_+)\,=\,0
\,\,\,\,\,\,\,\,\,\,\,\,
\Delta (\rho_-)\,=\,2
\,\,\,\,\,\,\,\,\,\,\,\,
\Delta (\lambda_+)\,=\,{1\over 2}
\,\,\,\,\,\,\,\,\,\,\,\,
\Delta (\lambda_-)\,=\,{3\over 2}.
\eqn\veinte
$$
Therefore $(\rho_-,\gamma^-)$ and $(\lambda_-,\eta^-)$ have
acquired the right dimensions to become the superdiffeomorphism
ghosts of the $N=1$ string. A glance at eqs. \dnueve\ and
\dsiete\ reveals that the other ghosts can be accommodated in
quartets with fields coming from the current sector. Indeed one
can pair $(\rho_0,\gamma^0)$ and $(\rho_+,\gamma^+)$ with
$(w^1,\chi^1)$ and $(w^2,\chi^2)$. The commuting ghosts
$(\lambda_+,\eta^+)$ can be paired with the
$({1\over 2},{1\over 2})$ fermionic system obtained from, say, the
fields ${1\over \sqrt{2}}\,(\psi^1+\bar\psi^1)$ and
${1\over \sqrt{2}}\,(\psi^2+\bar\psi^2)$.
After this process there remain two Majorana fields
$\psi_M={i\over \sqrt{2}}\,(\psi^1-\bar\psi^1)$ and
$\psi_L={i\over \sqrt{2}}\,(\psi^2-\bar\psi^2)$. Calling $\phi_M$,
$\phi_L$, $b$, $c$, $\beta$ and $\gamma$
 to $\phi_1$, $\phi_2$, $\rho_-$, $\gamma^-$, $\lambda_-$ and
$\eta^-$ respectively, we can write
the reduced energy-momentum tensor as:
$$
\eqalign{
T_{{\rm reduced}}\,=&\,
-{1\over 2}\,(\partial\phi_M)^2\,
-\,{1\over 2}\,(\partial\phi_L)^2\,+\,
{i\over 2}\,{t-1\over\sqrt t}\,\partial^2\phi_M\,-\,
{1\over 2}\,{t+1\over\sqrt t}\,\partial^2\phi_L\,-\cr
&-{1\over 2}\,\psi_M\partial\psi_M\,-\,
{1\over 2}\,\psi_L\partial\psi_L\,
+\,2b\partial c\,-\,c\partial b\,+\,{3\over 2}\partial\gamma
\beta\,+\,{1\over 2}\,\gamma\partial\beta,\cr}
\eqn\vuno
$$
which is indeed the one corresponding to the $N=1$ RNS
superstring. Furthermore the matter central charge in
$T_{{\rm reduced}}$ is:
$$
c_M\,=\,{3\over2}\,(\,1\,-\,2\,{(t-1)^2\over t}\,).
\eqn\vdos
$$
When $t=2k+3={q\over p}$ with $p,q\in \ZZ$, eq. \vdos\ gives the
central charge of the minimal models of the $N=1$ superconformal
symmetry.

To finish this section let us now analyse  the ${\rm
osp}(2|2)$ current algebra. This algebra contains four
bosonic currents  ($J_{\pm}$, $H$ and $J$) along with
other four fermionic ones ($j_{\pm\pm}$). The currents
$J_{\pm}$ and $H$ close an ${\rm sl}(2)$ algebra, while
$J$ is a $U(1)$ current. One can represent this algebra by
means of two scalar fields $\phi$ and $\varphi$, one
$(1,0)$ commuting $\beta\gamma$ system (denoted by
$(w,\chi)$) and two $(1,0)$ anticommuting $bc$ systems
($(\bar\psi_+,\psi_-)$ and $(\bar\psi_-,\psi_+)$). We
shall use the conventions of eq. \doce\ for the OPE's of
the fields $(w,\chi)$, $\phi$ and $\varphi$. For the
fermionic fields, the basic OPE's are:
$$
\psi_+(z)\bar\psi_{-}(w)\,=\,\psi_-(z)\bar\psi_+(w)\,=\,
{1\over z-w}.
\eqn\vtres
$$
Then the explicit representation[\bershadsky] of the ${\rm osp}(2|2)$
 currents is:
$$
\eqalign{
J_+\,=&\,w\cr
J_-\,=&-w\chi^2\,+\,i\sqrt{2k+2}\,\,\chi\partial\phi\,
-\chi(\,\psi_-\bar\psi_+\,+\,\psi_+\bar\psi_-\,)\cr
-&\sqrt{2k+2}\,\,\psi_+\psi_-\partial\varphi\,+\,
k\partial\chi\,+\,(k+1)[\,\psi_-\partial\psi_+\,+\,
\,\psi_+\partial\psi_-\,]\cr
H\,=&-w\chi\,+\,{i\over 2}\,\sqrt{2k+2}\,\,\partial\phi\,
-\,{1\over 2}\,
[\psi_-\bar\psi_+\,+\,\psi_+\bar\psi_-\,]\cr
J\,=&\,-\,{1\over 2}\,
[\psi_-\bar\psi_+\,-\,\psi_+\bar\psi_-\,]
\,+\,{\sqrt{2k+2}\over 2}\,\,\partial\varphi\cr
j_{+\pm}\,=&\,\bar\psi_{\pm}\,+\,w\psi_{\pm}\cr
j_{-\pm}\,=&\,-\chi\,(\bar\psi_{\pm}\,+\,w\psi_{\pm}\,)
\,+\,\sqrt{2k+2}\,\,\psi_{\pm}\, \partial(i\phi\,\pm\,
\varphi\,)\,+\,(2k+1)\,\partial\psi_{\pm}\,+\,
\psi_{\pm}\psi_{\mp}\bar\psi_{\pm}.\cr\cr}
\eqn\vcuatro
$$

A direct computation using the operator algebra closed by the
currents of eq. \vcuatro\ shows that $C_A=2$ for ${\rm osp}(2|2)$.
This same calculation yields the values of the metric tensor.
Using these values we can write down the
Sugawara tensor:
$$
\eqalign{
T^J\,=\,{1\over
2k+2}\,&[\,J_+J_-\,+\,J_-J_+\,+\, 2H^2\,-\,2J^2\,
-\,{1\over
2}\,(\,j_{++}\,j_{--}\,-\,\,j_{--}\,j_{++}\,+\,\cr
+&\,j_{+-}\,j_{-+}\,-\,\,j_{-+}\,j_{+-})\,].\cr}
\eqn\vcinco
$$
Taking eq.\vcuatro\ into account, one can obtain the expression of
$T^J\,\,$ in terms of the free fields. After some
calculation one gets:
$$
T^J\,=\,w\partial\chi\,-\,
\bar\psi_+\partial\psi_-\,-\,\bar\psi_-\partial\psi_+
\,-\,{1\over 2}\,(\partial \phi)^2\,-\,
{1\over 2}\,(\partial \varphi)^2.
\eqn\vseis
$$
Notice that the fields $\phi$ and $\varphi$ have vanishing
background charges. It is also evident by inspecting eq.
\vseis\ that the central charge of the ${\rm osp}(2|2)$ WZW model is
zero. This also follows from the fact that
$d_B\,=\,d_F\,=\,4$
for ${\rm osp}(2|2)$(see eq. \zk).

As in the ${\rm osp}(1|2)$ case, we shall denote the ghosts for the
currents $H$ and $J_{\pm}$ by $(\rho_0,\gamma^0)$ and
$(\rho_{\pm},\gamma^{\pm})$, whereas $(\rho_J,\gamma^J)$ and
$(\lambda_{\pm\pm},\eta^{\pm\pm})$ will correspond to the
currents $J$ and $j_{\pm\pm}$ respectively. The topological
${\rm osp}(2|2)$ current system will be realised by combining these
ghosts with two currents whose  levels are $k_1=k$ and $k_2=-k-2$. In
order to make contact with the non-critical $N=2$ superstring we
must first improve the energy-momentum tensor. Let us assume that
we deform the total operator $T$ as $T\rightarrow T+\partial{\cal H}$,
where ${\cal H}$ is the total Cartan current along the
$H$-direction. The contributions to ${\cal H}$ of the
currents with levels $k$ and $-k-2$ can be read from the
third equation in \vcuatro. Moreover,  using the structure
constants of the ${\rm osp}(2|2)$ algebra, it is easy to compute the ghost
contribution to ${\cal H}$. One gets:
$$
\eqalign{
{\cal H}^{{\rm gh}}\,=&\,
\,\gamma^+\rho_+\,-\,\gamma^-\rho_-\,-\,
{1\over 2}\,\eta^{++}\lambda_{++}\,-\,
{1\over 2}\,\eta^{+-}\lambda_{+-}\,+\,\cr\cr
+&\,{1\over 2}\,\eta^{-+}\lambda_{-+}\,+\,
{1\over 2}\,\eta^{--}\lambda_{--}.\cr}
\eqn\vsiete
$$
The improved energy-momentum tensor of the ghosts can be easily
computed from eq.\vsiete:
$$
\eqalign{
T&_{{\rm improved}}^{{\rm gh}}\,=
\,\rho_0\partial\gamma^0\,+\,
\,\rho_J\partial\gamma^J\,+\,
\gamma^+\partial\rho_+\,+\,2\rho_-\partial\gamma^- \,-\,
\gamma^-\partial\rho_- +\cr\cr
+&\sum_{\alpha = \pm}[
\,{1\over 2}\partial \eta^{+\alpha}\lambda_{+\alpha}
-\,{1\over 2} \eta^{+\alpha}\partial\lambda_{+\alpha}\,+\,
{3\over 2}\partial \eta^{-\alpha}\lambda_{-\alpha}
\,+\,{1\over 2}
\eta^{-\alpha}\partial\lambda_{-\alpha}\,].\cr}
\eqn\vocho
$$
Therefore the conformal  weights that the different antighosts
acquire after the deformation are:
$$
\eqalign{
&\Delta (\rho_0)\,=\,\Delta (\rho_J)\,=\,1
\,\,\,\,\,\,\,\,\,\,\,\,
\Delta (\rho_+)\,=\,0
\,\,\,\,\,\,\,\,\,\,\,\,
\Delta (\rho_-)\,=\,2\cr\cr
&\,\,\,\,\,\,\,\,\,\,\,\,\,\,\,\,\,\,\,\,\,
\Delta (\lambda_{+\pm})\,=\,{1\over 2}
\,\,\,\,\,\,\,\,\,\,\,\,
\Delta (\lambda_{-\pm})\,=\,{3\over 2}.\cr}
\eqn\vnueve
$$
Moreover, from eqs. \vseis\ and \vcuatro, one can obtain the improved
energy-momentum tensor in the current sector. If we label with the
indices $1$ and $2$ the free fields coming from the two currents, it is
straightforward to arrive at the result:
$$
\eqalign{
T&_{{\rm improved}}^{\rm J}\,=\,
\sum_{i=1,2}[\,
-{1\over 2}\,(\partial\varphi_i)^2\,
-\,{1\over 2}\,(\partial\phi_i)^2\,]\,+\,
{i\over 2}\,\sqrt{2k+2}\,\partial^2\phi_1\,-\,
{1\over 2}\,\sqrt{2k+2}\,\partial^2\phi_2\,-\cr\cr
-&\sum_{i=1,2}\,[\partial w^i\chi^i\,
+\, {1\over 2}\bar \psi^i_+\partial\psi^i_-\,
+\, {1\over 2}\psi^i_-\partial\bar\psi^i_+\,
+\, {1\over 2}\bar \psi^i_-\partial\psi^i_+\,
+\, {1\over 2}\psi^i_+\partial\bar\psi^i_-\,].\cr}
\eqn\treinta
$$

Let us now see how one can pair ghost fields from eq.\vocho\ with
fields in eq. \treinta\ in such a way that, after the reduction, we are
left with the field content of the non-critical $N=2$ superstring.
Indeed, as in the ${\rm osp}(1|2)$ case, we can pair the systems
$(\rho_0,\gamma^0)$ and $(\rho_+,\gamma^+)$ with
$(w^1,\chi^1)$ and $(w^2,\chi^2)$. Moreover we can form a quartet with
the ghosts $(\lambda_{\pm\pm}, \eta^{\pm\pm})$ and the
$({1\over 2},{1\over 2})$ anticommuting system formed from the Majorana
fermions ${1\over 2}(\psi^1_{\pm}+\bar \psi^1_{\mp})$ and
${1\over 2}(\psi^2_{\pm}+\bar \psi^2_{\mp})$. The remaining fields can
be assigned to matter and Liouville degrees of freedom. Let us first
consider the bosonic fields. If the labels M and L refer matter and
Liouville fields, we can define:
$$
\eqalign{
\phi_M\,=&\,{1\over\sqrt 2}\,(\phi_1\,+\,i\varphi_1)
\,\,\,\,\,\,\,\,\,\,\,\,\,\,\,\,\,
\bar\phi_M\,=\,{1\over\sqrt
2}\,(\phi_1\,-\,i\varphi_1)\cr\cr
\phi_L\,=&\,{1\over\sqrt 2}
\,(\phi_2\,+\,i\varphi_2)
\,\,\,\,\,\,\,\,\,\,\,\,\,\,\,\,\,
\bar\phi_L\,=\,{1\over\sqrt 2}\,
(\phi_2\,-\,i\varphi_2).\cr}
\eqn\tuno
$$
Similarly we can define the Dirac fermionic fields:
$$
\eqalign{
\psi_M\,=&\,{i\over 2}\,[\psi^1_+\,-\,\bar\psi^1_-\,+\,
i(\psi^1_-\,-\,\bar\psi^1_+)\,]
\,\,\,\,\,\,\,\,\,\,\,\,\,\,\,\,\,
\bar\psi_M\,=\,{i\over 2}\,[\psi^1_+\,-\,\bar\psi^1_-\,-\,
i(\psi^1_-\,-\,\bar\psi^1_+)\,]\cr\cr
\psi_L\,=&\,{i\over 2}\,[\psi^2_+\,-\,\bar\psi^2_-\,+\,
i(\psi^2_-\,-\,\bar\psi^2_+)\,]
\,\,\,\,\,\,\,\,\,\,\,\,\,\,\,\,\,
\bar\psi_L\,=\,{i\over 2}\,[\psi^2_+\,-\,\bar\psi^2_-\,-\,
i(\psi^2_-\,-\,\bar\psi^2_+)\,].\cr\cr}
\eqn\tdos
$$
It should be noticed that $\psi_M$ and $\bar\psi_M$
($\psi_L$ and $\bar\psi_L$) are the components of the fermionic fields
used to represent the ${\rm osp}(2|2)$ current at level $k_1$ (respectively
$k_2$) that are not affected by the reduction described above.
Splitting the reduced energy-momentum tensor as:
$$
T_{{\rm reduced}}\,=\,T_{{\rm reduced}}^{\rm M+L}\,+\,
T_{{\rm reduced}}^{\rm gh},
\eqn\ttres
$$
we can easily write down the matter + Liouville contributions.
Using the definitions \tuno\ and \tdos, one arrives at:
$$
\eqalign{
T_{{\rm reduced}}^{\rm M+L}\,=&\,
-\partial\phi_M\partial\bar\phi_M\,
-\partial\phi_L\partial\bar\phi_L\,+\,
{\sqrt{k+1}\over 2}\,
(i\partial^2\,(\phi_M+\bar\phi_M)\,-\,
\partial^2\,(\phi_L+\bar\phi_L))-\cr\cr
-&{1\over 2}\,\bar\psi_M\partial\psi_M
-{1\over 2}\,\psi_M\partial\bar\psi_M-
{1\over 2}\,\bar\psi_L\partial\psi_L
-{1\over 2}\,\psi_L\partial\bar\psi_L.\cr\cr}
\eqn\tcuatro
$$
The ghost part of $T_{{\rm reduced}}$ can be read form eq. \vocho.
Relabelling the ghost fields unaffected by the reduction as
$\rho_-\rightarrow b$, $\gamma^-\rightarrow c$,
$\rho_J\rightarrow \tilde b$, $\gamma^J\rightarrow \tilde c$,
$\lambda_{-\pm}\rightarrow \beta_{\pm}$ and
$\eta^{-\pm}\rightarrow \gamma^{\pm}$, one can finally write
$T_{{\rm reduced}}^{\rm gh}$ in the form:
$$
T_{{\rm reduced}}^{\rm gh}\,=\,
2b\partial c\,-\,c\partial b\,+
\,{3\over 2}\,\partial\gamma^+
\beta_+\,+\,{1\over 2}\,\gamma^+\partial\beta_+\,+\,
\,{3\over 2}\,\partial\gamma^-
\beta_-\,+\,{1\over 2}\,\gamma^-\partial\beta_-\,+\,
\tilde b\partial\tilde c,
\eqn\tcinco
$$
which indeed coincides with the energy-momentum tensor for the
ghosts of the $N=2$ string. Notice that the contribution of this $N=2$
ghost system to the Virasoro central charge is $-6$. Moreover the
matter central charge $c_M$ and the ${\rm osp}(2|2)$ level $k$ are now
related as:
$$
c_M\,=\,-6k-3,
\eqn\tseis
$$
in complete agreement with ref. [\bershadsky]

\chapter{Summary and conclusions.}

In this paper we have analysed the topological
 structure associated with the Ka\v c--Moody symmetry
based on an arbitrary Lie superalgebra ${\cal G}$. The resulting
 topological algebra turns out to
be an extension of the twisted  $N=2$ superconformal
algebra [\LVW,\EY] by a BRST doublet of spin 3
operators, upon whose introduction the algebra
closes linearly.
Our method consists in introducing a ghost sector that
allows to define a BRST cohomology, in such a
way that all operators  appear as BRST doublets.  The
resulting theory has been interpreted as a
topological sigma model for supergroup manifolds,
\ie, as a ${\cal G}/{\cal G}$ coset.

By means of  free field representations we have related the ${\rm osp}(1|2)$
and
${\rm osp}(2|2)$ models with the $N=1$ and $N=2$ non-critical superstrings,
respectively. We have established that, after performing a suitable
deformation of the coset model, one can identify the matter, Liouville
and ghost sectors of the coset with those of the
corresponding string theory. Moreover the
remaining degrees of freedom can be accommodated in topological doublets.

In this paper we have restricted ourselves to comparing the field content
of the deformed coset models and non-critical superstrings. It remains
to see whether or not the BRST cohomologies of both models are related.
 For bosonic
$W_N$-strings this relation has been shown to exist
[\yank,\hu,\sadov]
and thus one expects a similar result for the supersymmetric string
models (see ref. [\yu] for an analysis of the ${\rm osp}(1|2)$ case).
There are many other
questions that remain open in the relation between the
${\cal G}/ {\cal G}$ cosets and string theories. For example, one
would expect to have a current algebra prescription to compute
correlation functions in non-critical string theories. Another
problem that, in our opinion, deserves future investigation,  is the
relation between the ${\cal G}/{\cal G}$ BRST symmetry and the
topological symmetry of non-critical strings discovered in ref.
\REF\gato{B. Gato-Rivera and A.M.
Semikhatov\journal\pl&B293(92)72 \journal\np&B408(93)133.}
\REF\BLNW{M. Bershadsky, W. Lerche, D. Nemechansky and N.
Warner\journal\np&B401(93)304.}
[\gato] (see also [\BLNW]).

The ${\cal G}/{\cal G}$ coset models can also be regarded as
models of topological matter. The coupling of this
${\cal G}/ {\cal G}$ matter to topological gravity gives rise to
a topological string model. A prescription for this coupling was given
in ref. [\getzler] at the level of the operator algebra. The extended
character of the BRST algebra plays an important role in this analysis.
It would be interesting to understand this coupling within a lagrangian
formalism. This  could possibly shed light on the nature and
implications of the BRST symmetry of current algebras.

\ack
The authors would like to thank
J. M. F. Labastida and J. S\'anchez Guill\'en for encouragement.
Discussions with J. S\'anchez de Santos, P. M. Llatas
and L. Ferreira are gratefully acknowledged. J. M. I. is
thankful to the Theory Group of the Physics Department
of Brandeis University for hospitality,
 and to  Ministerio de
Educaci\'on y Ciencia (Spain) for financial support.
 This work was supported in part by DGICYT under
grant PB90-0772, and by CICYT under grants  AEN90-0035 and AEN93-0729.

\refout

\end